\documentstyle[11pt,paspconf,psfig]{article}

\newenvironment{bul}
{\begin{list}
  {$\quad \bullet$}
  {\itemsep = 0.5ex\parsep=0pt\topsep = 0.5mm}}
{\end{list} }

\begin{document}

\title{The emission line properties of the 3CR radio galaxies at redshift
one: shocks, evolution, and the alignment effect}

\author{Philip Best and Huub R\"ottgering}
\affil{Sterrewacht Leiden, Postbus 9513, 2300\,RA Leiden, The Netherlands}

\author{Malcolm Longair}
\affil{Cavendish Astrophysics, Madingley Road, Cambridge, CB3 0HE, UK}

\begin{abstract}
The results of a deep spectroscopic campaign on powerful radio galaxies
with redshift $z \sim 1$, to investigate in detail their emission line gas
properties, are presented. Both the 2-dimensional velocity structure of
the [OII]~3727 emission line and the ionisation state of the gas are found
to be strongly dependent upon the linear size (age) of the radio source in
a manner indicative of the emission line properties of small (young) radio
sources being dominated by the passage of the radio source shocks. The
consequences of this evolution throughout the few $\times 10^7$ year
lifetime of the radio source are discussed, particularly with relation to
the alignment of the UV--optical continuum emission of these objects along
their radio axis, the nature of which shows similar evolution.
\end{abstract}

% Keywords should be included, but they are not printed in the hardcopy.


\vspace*{-2mm}
\section{Introduction}

Powerful radio galaxies possess extremely luminous extended emission line
regions, often aligned along the radio axis (e.g. McCarthy et~al. 1996 and
references therein). The source of ionisation of this gas has been a long
standing question.  Robinson et~al. (1987) found that optical emission
line spectra of most low redshift ($z \la 0.1$) radio galaxies are well
explained using photoionisation models, and a similar result was found for
a composite spectrum of radio galaxies with redshifts $0.1 < z < 3$
(McCarthy 1993). Photoionisation models are supported by
orientation--based unification schemes of radio galaxies and radio--loud
quasars (e.g. Barthel 1989), in which the radio galaxies will host an
obscured quasar nucleus. On the other hand, detailed studies of individual
sources (e.g. 3C171; Clark et~al. 1998) have revealed features such as
enhanced nebular line emission, large velocity widths and ionisation state
minima coincident with the radio hotspots, indicating that the morphology,
kinematics and ionisation of the gas in some sources are dominated by
shocks.

Powerful radio galaxies with $z \ga 0.6$ also show enhanced optical--UV
emission, which is elongated and aligned along the radio axis in a manner
similar to the line emission. In recent years we have observed a sample of
28 3CR radio galaxies with $0.6 < z < 1.8$ using the HST, VLA and UKIRT,
to study this aligned emission (e.g. Best et~al. 1997). An important
result is that the nature of the alignment effect evolves with increasing
size (corresponding roughly to increasing age) of the radio source (Best
et~al. 1996). Small radio sources show intense strings of blue knots which
track the passage of the radio jets, whilst larger sources typically have
much more diffuse optical--UV emission. Clearly the passage of the radio
jet has an important influence on these host galaxies.

In this contribution, the first results of a program of deep spectroscopic
observations on this sample of distant radio galaxies are presented. The
observations are described in Section~2. The main results are presented in
Sections~3 \& 4, and in Section~5 the implications of these results for
both the emission line gas and the alignment effect of powerful radio
galaxies are discussed.

\vspace*{-2mm}
\section{Sample selection and observations}

From our complete HST sample of 28 $z \sim 1$ 3CR radio galaxies (Best
et~al 1997), our spectroscopic studies were restricted initially to those
18 galaxies with redshifts $0.7 < z < 1.25$. Of these, 3C41, 3C65, 3C267
and 3C277.2 were not observed due to constraints of telescope time. The
exclusion of these four galaxies was based upon their right ascensions,
not upon source properties, and so their exclusion should not introduce
any significant selection effects.

The remaining 14 galaxies were observed for between 1.5 and 2 hours each
during July 1997 and February 1998, using the twin-armed ISIS spectrograph
on the William Herschel Telescope.  Low ($\sim 12$\AA) resolution spectra
in the blue arm provided a useful observed--frame wavelength coverage of
$\sim 3200$ to $\sim 5200$\AA, and intermediate ($\sim 5$\AA) resolution
red arm spectra sampled the rest--frame wavelength range from $\sim
3500$\AA\ to $\sim 4300$\AA. This setup covered a broad range of emission
lines, allowing investigation of the ionisation state of the gas, and
provided reasonably high resolution data on the strong [OII]~3727 emission
line, enabling the velocity structures to be determined. Details of the
dataset and the data reduction techniques are given by Best et~al. (1999).

\vspace*{-2mm}
\section{Emission line ratios}

The line ratios of CIII]\,1909\,/\,CII]\,2326 and
[NeIII]\,3869\,/\,[NeV]\,3426 have been determined for these galaxies, the
former ratio from the ISIS blue arm data, and the latter from the red arm
data (where possible) or from the literature. These line ratios are
particularly useful for ionisation studies for three reasons: (i) in both
cases the two lines in the ratio involve the same element, and so
variations in metallicity or abundance are not important; (ii) the two
lines are very close in wavelength, and so differential extinction is
minimised; (iii) the theoretical predictions of photoionisation and
ionisation by shocks for these line ratios are very different (see below).

Theoretical predictions for these line ratios in photoionisation models
have been taken from the study of Allen et~al. (1998), who generated these
two line ratios (amongst others) using the MAPPINGS II code
(e.g. Sutherland et~al. 1993) for the simple model of a planar slab of
material being illuminated by power--law spectrum of ionising
radiation. For two different spectral indices of the input spectrum
($F_{\nu} \propto \nu^\alpha$ with $\alpha$\,=\,$-$1 and
$\alpha$\,=\,$-$1.4), and two different densities of cloud ($n_{\rm e} =
100$ and 1000\,cm$^{-3}$), the ratios were calculated for a wide range of
ionisation parameter $U$, defined as the ratio of ionising photons
striking the cloud to the density of the cloud $[U = (c n_{\rm H})^{-1}
\int_{\nu_0}^{\infty} (F_{\nu} {\rm d}\nu)/h\nu]$.

Line ratios for ionisation by shocks were also calculated, using the
models of Dopita and Sutherland (1996). These authors tabulated
theoretical line strengths for a variety of different shock velocities
(150 to 500\,km\,s$^{-1}$), and `magnetic parameters' ($0 \le B/\sqrt {n}
\le 4$\,$\mu$G\,cm$^{-1.5}$).  The line ratios were calculated both for the
shocked gas, and for the combination of shocked gas with a precursor
region, the latter corresponding to emission from the pre--shock gas due
to ionising photons produced by the shock diffusing upstream ahead of the
shock front (see Dopita and Sutherland 1996 for further discussion).

\begin{figure}
\psfig{figure=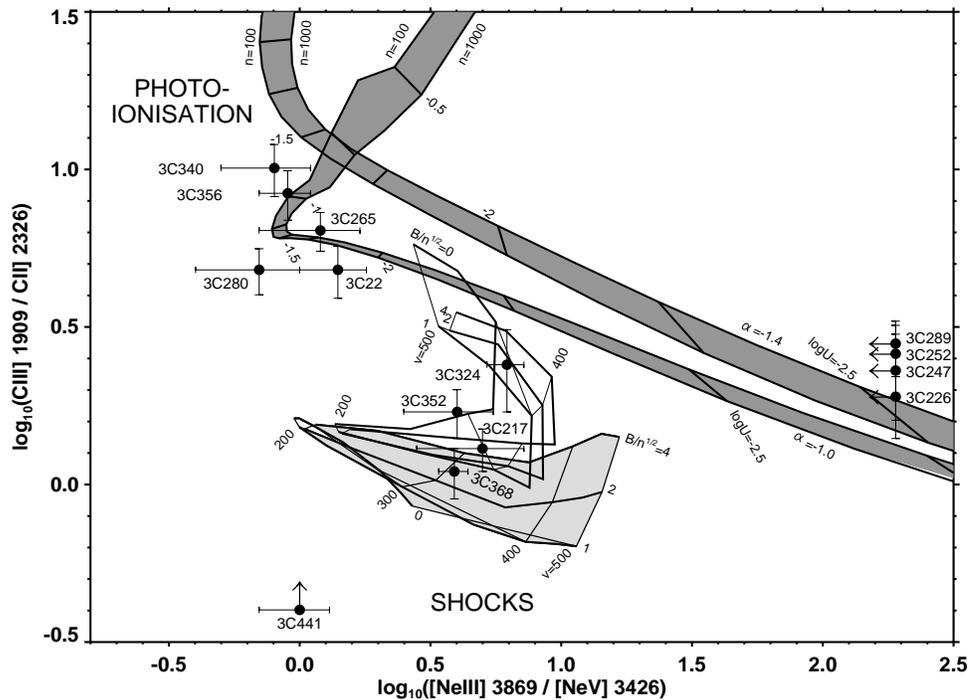,clip=,angle=90,width=\textwidth} 
\caption{An emission line diagnostic plot for the 3CR radio galaxies,
compared with theoretical predictions. The upper shaded regions correspond
to photoionisation models, the lower shaded region to shock models, and
the lower unshaded lines to shock models including a precursor region (see
text for details).}
\end{figure}

The output of these theoretical calculations are compared with the data in
Figure~1. Four galaxies (3C217, 3C324, 3C352, 3C368) lie in the region of
the diagram appropriate for shock ionisation, and five (3C22, 3C265,
3C280, 3C340, 3C356) lie close to the photoionisation models. The five
sources plotted at the edges of the plots have no data for one of their
emission lines. Interestingly, all of the four radio sources in the shock
region have radio sizes smaller than 115\,kpc ($\Omega = 1$, $H_0 =
50$\,km\,s$^{-1}$\,Mpc$^{-1}$), and the five `photoionised sources' have
larger radio sizes. Smaller radio sources appear to have lower ionisation
states.

\section{Velocity structures of the emission line gas}

The 2-dimensional [OII]~3727 velocity profiles of all 14 of the galaxies,
along with their intensity distributions, are presented in Best
et~al. (1999). The most important aspects of those profiles are as
follows. 

\begin{bul}
\item There is a strong inverse correlation between the FWHM of the
[OII]~3727 emission and the size of the radio source (Figure~2a). The four
`shock--dominated' sources from the previous section have the highest
FWHM.

\item Large radio sources often have smooth `rotation-dominated' velocity
profiles, whilst those of small (lower ionisation; see above) sources are
more distorted. Note that Baum et~al. (1992) found that radio galaxies
with redshifts $z < 0.2$ whose emission line velocities were consistent
with rotation typically had a high ionisation state
([OIII]\,5007\,/\,H$\beta \ga 5$); `non-rotators' had lower ionisation
states.

\item The [OII]~3727 line flux, normalised by the integrated K--band flux
which essentially measures the stellar mass of the galaxy, correlates
inversely with the radio source size. The four `shock--dominated' sources
all have high integrated [OII]~3727 fluxes (Figure~2b).
\end{bul}

\vspace*{-2mm}

\section{Discussion}

These results can be fit together in a scenario whereby the passage of the
radio bow shocks through the host galaxy dominates the kinematics and
ionisation of smaller (younger) radio sources, but large radio sources are
more relaxed and photoionisation from the AGN dominates. As the jet passes
through the emission line regions: (1) the emission line gas will be
accelerated by the shock, giving rise to the larger observed FWHM, and the
distorted [OII] velocity structures; (2) the shock will provide additional
ionising photons, increasing the [OII] line emission; (3) the emission
line clouds will be compressed by the shock, leading to a decrease in the
degree of ionisation of the gas.

\begin{figure}
\psfig{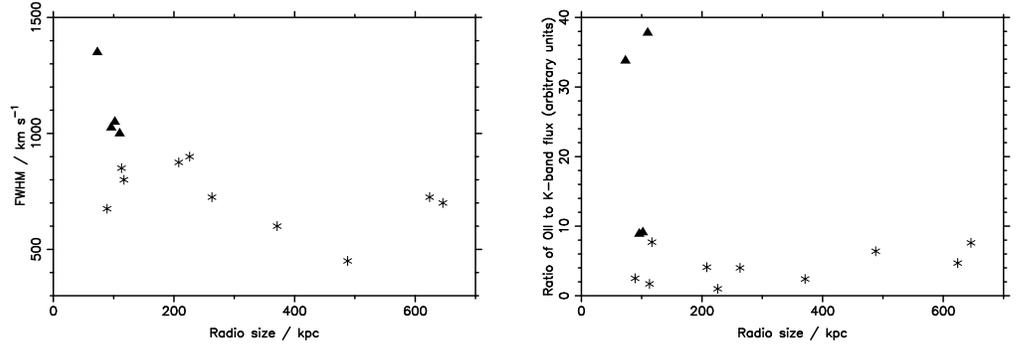} 
\caption{(a) The inverse correlation between the maximum FWHM of the
[OII]~3727 emission line and the radio source size; (b) The relationship
between the K--band (mass) normalised [OII] line intensity and radio
size. In each case, the four sources in the shock--dominated region of
Figure~1 are plotted as filled triangles, the remainder as asterisks.}
\end{figure}

Together with the results of Best et~al (1996,1998) that the aligned
optical--UV emission is tightly associated with the radio jet, these
results suggest that radio source shocks are also important for the
continuum alignment effect. Shock excitation of the emission--line clouds
will give rise to an enhanced contribution of nebular continuum emission
in small sources (e.g. see Dickson et~al 1995). Radio jet shocks may also
induce the formation of massive knots of bright young stars, which will
disperse and fade over the lifetime of the source. These two mechanisms
would each account for both the tight alignment of the optical--UV
continuum emission along the radio jet and the observed variation in the
luminosity of this emission with radio size. Moreover, shocks may also be
responsible for disrupting optically thick clouds along the radio jet
direction and exposing previously hidden dust grains (Bremer et~al 1997);
this would give rise to an enhanced contribution of scattered light,
distributed in a non--biconical manner.

In conclusion, these results show that the passage of the radio induced
shocks through the host galaxy of powerful radio sources plays a key role
in producing the emission line gas properties of these sources. Much of
the continuum alignment effect may have its origin in these same shocks.

\acknowledgments

The WHT is operated on the island of La Palma by the Isaac Newton Group in
the Spanish Observatorio del Roches de los Muchachos of the Instituto de
Astrofisica de Canarias. This work was supported in part by the Formation
and Evolution of Galaxies network set up by the European Commission under
contract ERB FMRX-- CT96--086 of its TMR programme. We are grateful to
Mark Allen for supplying the output of the MAPPINGS II code in digitised
form, and to Matt Lehnert for useful discussions.

\end{document}